\documentclass[a4paper,12pt]{article}
\usepackage{amsmath,amssymb}

\begin{document}
\setcounter{page}{0}

\begin{titlepage}

\bigskip

\begin{center}
{\LARGE\bf
Low-Energy Dynamics of\\ Noncommutative $CP^1$ Solitons\\ \vspace{3mm}
in 2+1 Dimensions}

\vspace{10mm}

\bigskip
{\renewcommand{\thefootnote}{\fnsymbol{footnote}}
\large\bf 
Ko Furuta\footnote{
E-mail: furuta@phys.chuo-u.ac.jp}, 
Takeo Inami, 
Hiroaki Nakajima\footnote{
E-mail: nakajima@phys.chuo-u.ac.jp}\\
and Masayoshi Yamamoto\footnote{
E-mail: yamamoto@phys.chuo-u.ac.jp}
}

\setcounter{footnote}{0}
\bigskip

{\small \it
Department of Physics, Chuo University\\
Kasuga, Bunkyo-ku, Tokyo 112-8551, Japan.\\
}

\end{center}
\bigskip


\begin{abstract}
We investigate the low-energy dynamics of the BPS solitons of the noncommutative $CP^1$ model 
in 2+1 dimensions using the moduli space metric of the BPS solitons. 
We show that the dynamics of a single soliton coincides with that in the commutative model. 
We find that the singularity in the two-soliton moduli space, which exists in the 
commutative $CP^1$ model, disappears in the noncommutative model.
We also show that the two-soliton metric has the smooth commutative limit.
\end{abstract}

\end{titlepage}


\section{Introduction}

Noncommutative geometry appears in M-theory, string theory and condensed matter physics 
\cite{Nek}.
Noncommutative field theories are known to describe the low-energy effective theory of D-branes
in a background B-field \cite{CDS,CK,SW}.
(2+1)-dimensional noncommutative theories have applications to the quantum Hall effect 
\cite{Suss etc}. 
In spite of the nonlocality and entangled UV/IR mixing 
in perturbation theory, the appearance of these theories 
in string theory suggests that some class of noncommutative 
field theories is a sensible deformation of ordinary 
field theories.

In view of understanding nonperturbative effects in noncommutative field theories,
noncommutative solitons and instantons have been investigated. 
In four-dimensional noncommutative Yang-Mills theory,
there exists a U(1) instanton which is nonsingular 
due to the position-space uncertainty \cite{NS}. The moduli 
space of instantons are also smooth \cite{LTY,Nak}. 
In noncommutative scalar field theories,
there exist solitons (GMS solitons) \cite{GMS}
at the limit of large noncommutativity parameter $\theta$;
they cannot exist in the commutative counterpart.

Low-energy dynamics of solitons can be approximated by
geodesic motion on the moduli space of static solutions \cite{Manton}.
In commutative theories,
scattering properties of solitons in gauge theories and nonlinear sigma models
were studied extensively using this approximation \cite{AH,Ward,Leese}.
A common feature is that 
the scattering of two solitons occurs at right angle for the head-on collision.
In noncommutative theories, 
the scatterings of solitons in 
the GMS model, Yang-Mills theories and integrable models
were investigated \cite{Gopak etc,HIO,LP}.

In this letter
we investigate the low-energy dynamics of
BPS solitons of the noncommutative $CP^1$ model in 2+1 dimensions.
It was shown that the noncommutative $CP^N$ model has the BPS solutions 
as the commutative model does \cite{LLY}. 
In the commutative $CP^N$ model,
the moduli space of solitons is known to be a K\"ahlar manifold \cite{Ruback}.
We show that the moduli space of the noncommutative $CP^N$ model is a K\"ahlar manifold too.
We calculate the $\theta$-dependence of the moduli space metric in the $CP^1$ case.

This letter is organized as follows.
In section 2,
we review the BPS solutions of the noncommutative $CP^N$ model.
In section 3,
we give the compact form of the K\"ahlar potential 
of the moduli space of these solutions.
In section 4,
we investigate the one-soliton metric of the noncommutative $CP^1$ model
and show that the motion of a single soliton is the same as that of the commutative model.
In section 5,
we study the two-soliton metric of the 
model and find that the singularity which exists in the commutative 
model disappears in the noncommutative model.
Furthermore,
we show that the two-soliton metric has the smooth commutative limit.

\section{BPS solution of the noncommutative $CP^N$ model}
We recapitulate the (2+1)-dimensional noncommutative $CP^N$ model, closely
following Lee, Lee and Yang \cite{LLY}.

We consider the (2+1)-dimensional field theory on the noncommutative space.
The noncommutativity is introduced by
\begin{equation}
[x,y]=i\theta, \quad \theta>0. \label{nc1}
\end{equation}
We set
\begin{equation}
z=\frac{1}{\sqrt{2}}(x+iy),\quad \bar{z}=\frac{1}{\sqrt{2}}(x-iy).\label{z}
\end{equation}
Then the equation \eqref{nc1} becomes
\begin{equation}
[z,\bar{z}]=\theta,\label{nc2}
\end{equation}
or
\begin{equation}
[a,a^\dag]=1,\quad a=\frac{z}{\sqrt{\theta}},\quad a^\dag=\frac{\bar{z}}{\sqrt{\theta}}.
\end{equation}
This is the algebra of the creation and annihilation operators. We use the Fock space of 
the quantum harmonic oscillator for the representation of the algebra \eqref{nc1}.

The derivative and the integral on the noncommutative space are given by
\begin{gather}
\partial_x\Phi=i\theta^{-1}[y,\Phi],\quad \partial_y\Phi=-i\theta^{-1}[x,\Phi],\\
\int d^2 x\mathcal{O}\rightarrow\mbox{Tr}\mathcal{O}=2\pi\theta\sum_{n\ge 0}\langle n|\mathcal{O}|n\rangle,
\end{gather}
where $|n\rangle\ (n=0,1,2,\cdots)$ are the basis of the Fock space.

The Lagrangian of the noncommutative $CP^N$ model \cite{LLY} is defined by 
\begin{align}
L&=\mbox{Tr}[D_\mu\Phi^\dag D^\mu\Phi+\lambda(\Phi^\dag \Phi-1)],\label{L}\\
D_\mu&=\partial_\mu\Phi-i\Phi A_\mu, \quad A_\mu=-i\Phi^\dag\partial_\mu\Phi,
\end{align}
where the field $\Phi={}^{t}(\phi_1,\phi_2,\ldots,\phi_{N+1})$ is the complex 
$(N+1)$-component vector, and $\lambda$ is the Lagrange multiplier field which gives the constraint $\Phi^\dag \Phi=1$.
This theory has the global $SU(N+1)$ symmetry and $U(1)$ gauge symmetry $\Phi(x)\rightarrow \Phi(x)g(x)$, 
where $g(x)\in U(1)$.

The energy functional is
\begin{equation}
E=\mbox{Tr}(|D_0 \Phi|^2+|D_z \Phi|^2+|D_{\bar{z}} \Phi|^2).
\end{equation}
We have the Bogomolnyi bound
\begin{equation}
E\ge\mbox{Tr}(|D_0 \Phi|^2)+2\pi|Q|,
\end{equation}
where $Q$ is the topological charge,
\begin{equation}
Q=\frac{1}{2\pi}\mbox{Tr}(|D_z \Phi|^2-|D_{\bar{z}} \Phi|^2).\label{Q}
\end{equation}
The BPS equation of the static solution is
\begin{align}
D_{\bar{z}}\Phi&=0\quad \mbox{for self-dual solution},\label{SD}\\
D_z\Phi&=0\quad \mbox{for anti-self-dual solution}.
\end{align}

We parametrize $\Phi$ as
\begin{equation}
\Phi=W(W^\dag W)^{-1/2},
\end{equation}
where $W$ is the complex $(N+1)$-component vector. We define the projection operator $P$ by
\begin{equation}
P=1-W(W^\dag W)^{-1}W^\dag.
\end{equation}
Then \eqref{L} and \eqref{Q} are
\begin{align}
L&=\mbox{Tr}\left[\frac{1}{\sqrt{W^\dag W}}\partial_\mu W^\dag P\partial^\mu W\frac{1}{\sqrt{W^\dag W}}\right],
\label{WCPNLag}\\
Q&=\frac{1}{2\pi}\mbox{Tr}\left[\frac{1}{\sqrt{W^\dag W}}(\partial_{\bar{z}}W^\dag P\partial_z W-
\partial_z W^\dag P\partial_{\bar{z}}W)\frac{1}{\sqrt{W^\dag W}}\right].
\end{align}
The BPS equation \eqref{SD} becomes
\begin{equation}
D_{\bar{z}}\Phi=P(\partial_{\bar{z}}W)(W^\dag W)^{-1/2}=0.\label{SD2}
\end{equation}
This equation is equivalent to $\partial_{\bar{z}}W=WV$, where $V$ is an arbitrary scalar. 
In the commutative case, using the $N$-component vector $w$, we set
$W={}^{t}(w,1)$ by the gauge transformation 
\begin{equation}
W \to W'=W \Delta(z,\bar{z})\label{gaugeDelta},
\end{equation}
where $\Delta(z,\bar{z})$ is an arbitrary scalar.
Then \eqref{SD2} becomes $\partial_{\bar{z}}w=0$, namely, $w$ is holomorphic.
In the noncommutative case, The Lagrangian \eqref{WCPNLag} is invariant 
under the transformation \eqref{gaugeDelta} when $\Delta$ is invertible.

We will be mainly concerned with the one- and two-soliton solutions 
of the noncommutative $CP^1$ model.
The one- and two-soliton solutions of the commutative $CP^1$ model are respectively given by 
\cite{Ward,Leese}
\begin{align}
w&=\lambda+\frac{\mu}{z-\nu},\label{c1}\\
w&=\alpha+\frac{2\beta z+\gamma}{z^2+\delta z+\epsilon},\label{c2}
\end{align}
where $\alpha,\beta,\cdots \in \boldsymbol{C}$ are the moduli parameters.
We may set the moduli parameters $\lambda$ and $\alpha$ to zero, using the global 
$SU(2)$ symmetry. 

In the noncommutative case, $W=(z,1)$ and $W'=Wz^{-1}=(1,z^{-1})$ are gauge inequivalent, 
where $z^{-1}$ is defined to be 
$z^{-1}\equiv(\bar{z}z)^{-1}\bar{z}\equiv\bar{z}(\bar{z}z+\theta)^{-1}$.
$z^{-1}$ satisfies $z z^{-1}=1$ and $z^{-1} z = 1- |0\rangle \langle 0|$.
$W$ satisfies the BPS equation \eqref{SD2} but $W'$ does not. 
In general, $W$ satisfies the BPS equation \eqref{SD2} 
if the components of $W$ are polynomials of $z$.
The BPS one- and two-soliton solutions in the noncommutative $CP^1$ model
corresponding to \eqref{c1} and \eqref{c2} are respectively given by
\begin{align}
W&=\binom{z-\nu}{\mu},\label{W1}\\
W&=\binom{z^2+\delta z+\epsilon}{2\beta z+\gamma}.
\end{align}

\section{Moduli space metric}
In the following section, we consider the scattering of BPS solutions of the noncommutative $CP^N$ model.
In the low-energy limit (near the Bogomolnyi bound), it is a good approximation that only the moduli parameters 
depend on the time \cite{Manton}. Their time evolution is determined by minimizing the action $S=\int\!dt\,L$.
It amounts to dealing with the kinetic energy $T$ (the term in the Lagrangian including the time derivative only), 
since the rest of the Lagrangian gives the topological charge.
The kinetic energy is given by
\begin{equation}
\begin{split}\label{T}
T&=\mbox{Tr}\left(\frac{1}{\sqrt{W^\dag W}}\partial_t W^\dag P\partial_t W\frac{1}{\sqrt{W^\dag W}}\right)\\
&=\frac{1}{2}\mbox{Tr}(\partial_t P')^2,
\end{split}
\end{equation}
where $P'$ is defined by
\begin{equation}
P'=1-P=W(W^\dag W)^{-1}W^\dag.
\end{equation}
$T$ is written as $T=\frac{1}{2}(ds/dt)^2$, where $ds^2$ is the line element of the moduli space. 
The dynamics of solitons is given by the geodesic line in the moduli space.
We denote generically the moduli parameters by $\zeta^a$. We then have
\begin{equation}
T=\frac{1}{2}g_{ab}\frac{d\zeta^a}{dt}\frac{d\zeta^b}{dt},
\end{equation}
where $g_{ab}$ is the moduli space metric.

It is convenient to express $P'$ in the Fock space language,
\begin{gather}
P'=\sum_{n,m}|\psi_n\rangle h^{nm} \langle\psi_m|,\quad |\psi_n\rangle=W|n\rangle,\\
h_{nm}=\langle\psi_n|\psi_m\rangle,\quad h^{nm}=(h_{nm})^{-1}.
\end{gather}
The BPS solution $W$ (or $|\psi_n\rangle$) is a holomorphic function of the moduli parameters.
It was shown that the moduli space in the case where $|\psi_n\rangle$ is holomorphic 
is a K\"ahler manifold \cite{GMS}. Hence, we write
\begin{equation}
T=\frac{1}{2}g_{\bar{a}b}\frac{d\zeta^{\bar{a}}}{dt}\frac{d\zeta^b}{dt},
\quad g_{\bar{a}b}=\frac{\partial}{\partial\zeta^{\bar{a}}}\frac{\partial}{\partial\zeta^b}K,
\end{equation}
where the K\"ahler potential $K$ is given by
\begin{equation}
K=\mbox{Tr}\ln(h_{nm})=\mbox{Tr}\ln(W^\dag W)\label{K}.
\end{equation}

\section{One-soliton metric}
The BPS one-soliton solution of the noncommutative $CP^1$ model is given by \eqref{W1}.
We substitute \eqref{W1} into \eqref{T}, then we have
\begin{equation}
\begin{split}\label{T1}
T&=\mbox{Tr}
\Bigg[
\frac{1}{\sqrt{(\bar{z}-\bar{\nu})(z-\nu)+|\mu|^2}}\partial_t 
\begin{pmatrix} \bar{z}-\bar{\nu} & \bar{\mu} \end{pmatrix}\\
&\times\left\{1-
\begin{pmatrix} z-\nu \\ \mu \end{pmatrix}
\frac{1}{(\bar{z}-\bar{\nu})(z-\nu)+|\mu|^2}
\begin{pmatrix} \bar{z}-\bar{\nu} & \bar{\mu} \end{pmatrix}
\right\}\\
&\times\partial_t
\begin{pmatrix} z-\nu \\ \mu \end{pmatrix}
\frac{1}{\sqrt{(\bar{z}-\bar{\nu})(z-\nu)+|\mu|^2}}
\Bigg].
\end{split}
\end{equation}
The $\Dot{\Bar{\mu}}\dot{\mu}$ term in $T$ is
\begin{equation}
2\pi\theta\Dot{\Bar{\mu}}\dot{\mu} \sum_{n\ge 0}\frac{1}{\theta n+|\mu|^2}\left[\frac{\theta n}{\theta n+|\mu|^2}
+\frac{|\nu|^2}{\theta(n+1)+|\mu|^2}\right].\label{div}
\end{equation}
The first term in \eqref{div} diverges. We set $\mu$ to a constant in the low-energy approximation.
Calculating the trace in \eqref{T1} with $\dot{\mu}=0$, we obtain
\begin{equation}
T=2\pi\frac{d\bar{\nu}}{dt}\frac{d\nu}{dt},
 \quad \text{or} \quad ds^2=4\pi d\bar{\nu}d\nu.\label{metric1}
\end{equation}
This is $\theta$-independent and coincides with the commutative case. A single soliton moves 
straight without changing the size.

The result \eqref{metric1} can be explained in a more general framework.
In the low-energy limit, the action $S=\int\!dt\,L$ is invariant 
under the Galilean transformation
\begin{align}
t&\to t\notag\\
x&\to x+v_x t\\
y&\to y+v_y t\ ,\notag
\end{align}
since this transformation does not change the commutation 
relation \eqref{nc1}. Under the Galilean transformation, 
the kinetic energy in the center-of-mass frame 
$T_{{\rm cm}}$ is transformed as
\begin{equation}
T_{{\rm cm}}\to \frac{1}{2}Mv^2+T_{{\rm cm}} \ ,\label{kineticE}
\end{equation}
where $M$ is the total mass of the solitons; $M=2\pi |Q|$. From \eqref{kineticE}, 
it follows that the contribution of the 
center-of-mass coordinates to the kinetic energy is the same as in 
the commutative case. From now on, we restrict the moduli 
parameters to the center-of-mass frame.

\section{Two-soliton metric}
The BPS two-soliton solution of the noncommutative $CP^1$ model in the center-of-mass frame 
($i.e.\ \delta=0$.) is
\begin{equation}
W=\binom{z^2+\epsilon}{2\beta z+\gamma}.\label{W2}
\end{equation}
Computing the kinetic energy in the low-energy limit, we can see that
the contribution of the $\Dot{\Bar{\beta}}\dot{\beta}$
term to the kinetic energy diverges. In the low-energy approximation, 
we set $\beta$ to a constant.
We consider the case of $\beta=0$ for simplicity. 
In the commutative model, the moduli space metric is calculated 
by Ward \cite{Ward}. There exists a singularity at $(\epsilon, \gamma)
=(0,0)$. 
In the next subsection, 
we will see the disappearance of this singularity in the 
noncommutative model.

The K\"ahler potential corresponding to \eqref{W2} is
\begin{equation}
K=\mbox{Tr}\ln\left[(\bar{z}^2+\bar{\epsilon})
(z^2+\epsilon)+\bar{\gamma}\gamma\right]\ ,\label{Kaehler}
\end{equation}
This is a formal expression since the trace in 
\eqref{Kaehler} diverges.
A finite K\"ahler potential is obtained by subtracting the divergent terms 
using the K\"ahler transformation
\begin{equation}
K(\gamma,\epsilon;\bar{\gamma},\bar{\epsilon}) \rightarrow K(\gamma,\epsilon;\bar{\gamma},\bar{\epsilon})
-f(\gamma,\epsilon)-\bar{f}(\bar{\gamma},\bar{\epsilon})
\end{equation}
In \eqref{Kaehler}, only the terms with the same number of $z^2$ 
and $\bar{z}^2$ contribute to the trace. 
Hence, $K$ is the function of $\bar{\epsilon}\epsilon$ 
and $\bar{\gamma}\gamma$ only; $K=K(\bar{\epsilon}\epsilon, \bar{\gamma}
\gamma)$. It then follows that the moduli 
space is manifestly invariant under the following
three kinds of transformations: 
i) $(\epsilon, \gamma)\leftrightarrow (\bar{\epsilon}, \bar{\gamma})$, 
ii) $\epsilon \rightarrow e^{i\phi}\epsilon$, 
and iii) $\gamma \rightarrow e^{i\chi}\gamma$.

The moduli space metric is
\begin{subequations}\label{metric2}
\begin{align}
g_{\bar{\gamma}\gamma}&=\mbox{Tr}\left[\frac{1}{\bar{\gamma}\gamma+(\bar{z}^2+\bar{\epsilon})(z^2+\epsilon)}
\left(1-\frac{\bar{\gamma}\gamma}{\bar{\gamma}\gamma+(\bar{z}^2+\bar{\epsilon})(z^2+\epsilon)}\right)\right],\label{gg}\\
g_{\bar{\epsilon}\gamma}&=-\mbox{Tr}\left[\bar{\gamma}(z^2+\epsilon)
\frac{1}{[\bar{\gamma}\gamma+(\bar{z}^2+\bar{\epsilon})(z^2+\epsilon)]^2}\right],\label{ge}\\
g_{\bar{\gamma}\epsilon}&=-\mbox{Tr}\left[\gamma\frac{1}{[\bar{\gamma}\gamma+
(\bar{z}^2+\bar{\epsilon})(z^2+\epsilon)]^2}(\bar{z}^2+\bar{\epsilon})\right],\label{eg}\\
g_{\bar{\epsilon}\epsilon}&=\mbox{Tr}\left[\frac{1}{\bar{\gamma}\gamma+(\bar{z}^2+\bar{\epsilon})(z^2+\epsilon)}
\frac{\bar{\gamma}\gamma}{\bar{\gamma}\gamma+(\bar{z}^2+\bar{\epsilon})(z^2+\epsilon)+4\theta\bar{z}z+2\theta^2}\right].
\label{ee}
\end{align}
\end{subequations}
It is difficult to compute the trace of \eqref{metric2} exactly, but we can investigate the moduli space metric 
of the two solitons in the case of $|\gamma|,|\epsilon| \ll \theta$ or $|\gamma|,|\epsilon| \gg \theta$.

\subsection{The case of $|\gamma|,|\epsilon| \ll \theta$}
In the case of $|\gamma|,|\epsilon| \ll \theta$, we write \eqref{gg} as
\begin{multline}
g_{\bar{\gamma}\gamma}=\frac{1}{\theta^2}\mbox{Tr}\Biggl[\frac{1}{\bar{\gamma}\gamma/\theta^2+({a^\dag}^2+
\bar{\epsilon}/\theta)(a^2+\epsilon/\theta)}\\ \times\left(1-\frac{\bar{\gamma}\gamma/\theta^2}{\bar{\gamma}\gamma/\theta^2
+({a^\dag}^2+\bar{\epsilon}/\theta)(a^2+\epsilon/\theta)}\right)\Biggr], \label{metric2L}
\end{multline}
The operator $a^2+\epsilon/\theta$ has two zero eigenstates 
$e^{\pm i\sqrt{\frac{\epsilon}{\theta}}a^\dag}|0\rangle$. These states do not contribute to the
trace in \eqref{metric2L}. Then, we can easily compute the trace in the lowest order of 
$\theta^{-1}$. For \eqref{ge}, \eqref{eg} and \eqref{ee}, we can calculate the trace similarly.
Then we obtain
\begin{equation}
ds^2=\frac{2\pi}{\theta}\left(d\bar{\gamma}d\gamma+\frac{2}{3}d\bar{\epsilon}d\epsilon\right)+O(\theta^{-2})\label{large}.
\end{equation}
Therefore, the metric is flat. 
The singularity which exists in the commutative model at 
$(\epsilon, \gamma)=(0,0)$ disappears in the noncommutative 
model. The same phenomena are known in noncommutative 
Yang-Mills theories \cite{LTY,Nak}.
Since the relative coordinate of the solitons is $i\epsilon^{1/2}$, 
it seems that the geodesic which connects $\epsilon=\epsilon_0$ 
and $\epsilon=-\epsilon_0$ represents the right angle scattering. 
However, \eqref{large} is valid only in the region $|\epsilon| \ll \theta$. 
To investigate the scattering, a further study of the moduli space is
needed.

\subsection{The case of $|\gamma|,|\epsilon| \gg \theta$}
In the case of $|\gamma|,|\epsilon| \gg \theta$, it is convenient to use the $\star$-product formalism to compute 
\eqref{metric2} rather than the operator formalism. The $\star$-product is defined by
\begin{equation}
f(z,\bar{z})\star g(z,\bar{z})=f(z,\bar{z})\exp\left[\frac{\theta}{2}
(\overleftarrow{\partial_z}\overrightarrow{\partial_{\bar{z}}}
-\overleftarrow{\partial_{\bar{z}}}\overrightarrow{\partial_z})\right]g(z,\bar{z}).
\end{equation}
We use the formulae
\begin{align}
\frac{1}{\bar{\gamma}\gamma+(\bar{z}^2+\bar{\epsilon})(z^2+\epsilon)}&=
\int_0^\infty du\exp\bigl[-u\{\bar{\gamma}\gamma+(\bar{z}^2+\bar{\epsilon})(z^2+\epsilon)\}\bigr]\notag\\
\rightarrow&\int_0^\infty du\exp_\star\bigl[-u\{\bar{\gamma}\gamma+(\bar{z}^2+\bar{\epsilon})\star(z^2+\epsilon)\}\bigr],
\label{star}\\
\frac{1}{[\bar{\gamma}\gamma+(\bar{z}^2+\bar{\epsilon})(z^2+\epsilon)]^2}&=
\int_0^\infty du u\exp\bigl[-u\{\bar{\gamma}\gamma+(\bar{z}^2+\bar{\epsilon})(z^2+\epsilon)\}\bigr]\notag\\
\rightarrow&\int_0^\infty du u\exp_\star\bigl[-u\{\bar{\gamma}\gamma+(\bar{z}^2+\bar{\epsilon})\star(z^2+\epsilon)\}\bigr],
\end{align}
where $\exp_\star$ is defined by
\begin{equation}
\exp_\star(A)=1+A+\frac{1}{2!}A\star A+\frac{1}{3!}A\star A\star A+\cdots.
\end{equation}
For small $\theta$, we have the following relation
\begin{equation}
\exp_\star(A)=\exp(A)+O(\theta^2).\label{exp}
\end{equation}
Using the formulae \eqref{star}-\eqref{exp}, the moduli space metric up to the order $\theta$ 
is
\begin{subequations}\label{metric2S}
\begin{align}
g_{\bar{\gamma}\gamma}&=\int d^2x\Biggl[\frac{1}{\bar{\gamma}\gamma+(\bar{z}^2+\bar{\epsilon})(z^2+\epsilon)}
\left(1-\frac{\bar{\gamma}\gamma}{\bar{\gamma}\gamma+(\bar{z}^2+\bar{\epsilon})(z^2+\epsilon)}\right)\notag\\
{}+{}&\theta\frac{2\bar{z}z}{[\bar{\gamma}\gamma+(\bar{z}^2+\bar{\epsilon})(z^2+\epsilon)]^2}\left(1
-\frac{2\bar{\gamma}\gamma}{\bar{\gamma}\gamma+(\bar{z}^2+\bar{\epsilon})(z^2+\epsilon)}\right)\Biggr]+O(\theta^2),\\
g_{\bar{\epsilon}\gamma}&=-\int d^2x\Biggl[\frac{\bar{\gamma}(z^2+\epsilon)}{[\bar{\gamma}\gamma+(\bar{z}^2
+\bar{\epsilon})(z^2+\epsilon)]^2}\notag\\ &\hspace{4.5cm}
+\theta\frac{4\bar{\gamma}\bar{z}z(z^2+\epsilon)}{[\bar{\gamma}\gamma
+(\bar{z}^2+\bar{\epsilon})(z^2+\epsilon)]^3}\Biggr]+O(\theta^2),\\
g_{\bar{\gamma}\epsilon}&=-\int d^2x\Biggl[\frac{\gamma(\bar{z}^2+\bar{\epsilon})}{[\bar{\gamma}\gamma+(\bar{z}^2
+\bar{\epsilon})(z^2+\epsilon)]^2}\notag\\ &\hspace{4.5cm}
+\theta\frac{4\gamma\bar{z}z(\bar{z}^2+\bar{\epsilon})}{[\bar{\gamma}\gamma
+(\bar{z}^2+\bar{\epsilon})(z^2+\epsilon)]^3}\Biggr]+O(\theta^2),\\
g_{\bar{\epsilon}\epsilon}&=\int d^2x\frac{\bar{\gamma}\gamma}{[\bar{\gamma}\gamma
+(\bar{z}^2+\bar{\epsilon})(z^2+\epsilon)]^2}+O(\theta^2).
\end{align}
\end{subequations}
The integrals appearing in the coefficients of $\theta$ in \eqref{metric2S} converge.
Hence, the moduli space metric has the smooth commutative limit $\theta\rightarrow 0$ with $\gamma$ and $\epsilon$ fixed.


\section*{Acknowledgements}
We would like to thank Ryu Sasaki and Katsushi Ito 
for valuable comments. 
K. F. and H. N. are supported by a Research 
Assistantship of Chuo University. This work is supported 
partially by grants of Ministry of Education, 
Science and Technology, (Priority Area B, ``Supersymmetry and 
Unified Theory" and Basic Research C). 


\end{document}